\documentclass[preprint,12pt]{elsarticle}
\usepackage{array}
\usepackage{graphicx}

\makeatletter
\newcommand{\figcaption}{\def\@captype{figure}\caption}
\newcommand{\tabcaption}{\def\@captype{table}\caption}
\makeatother
\usepackage{amssymb}
\newcounter{bla}

\journal{Computer Physics Communications}

\begin{document}

\begin{frontmatter}

\title{On the lepton-nucleon neutral and charged current deep inelastic
       scattering cross sections}

\author[ciae]{Xing-Long Li\corref{cor1}}
\ead{lixinglong.c@gmail.com}
\author[ciae]{Yu-Liang Yan}
\ead{yanyl@ciae.ac.cn}
\author[ciae]{Xiao-Mei Li}
\author[ccnu]{Dai-Mei Zhou}
\author[ccnu]{Xu Cai}
\author[ciae,ccnu]{Ben-Hao Sa\corref{cor1}}
\ead{sabh@ciae.ac.cn}

\cortext[cor1]{Corresponding author}

\address[ciae]{China Institute of Atomic Energy, P. O. Box 275 (10), Beijing,
               102413 China.}
\address[ccnu]{Key Laboratory of Quark and Lepton Physics (MOE) and Institute
  of Particle Physics, Central China Normal University, Wuhan 430079, China.}

\begin{abstract}
Based on the requirement in the simulation of lepton-nucleus deep inelastic
scattering (DIS), we construct a fortran program LDCS 1.0 calculating the
differential and total cross sections for the unpolarized charged lepton-
unpolarized nucleon and neutrino-unpolarized nucleon neutral current
(charged current) DIS at leading order. Any set of the experimentally fitted
parton distribution functions could be employed directly. The mass of incident
and scattered leptons is taken into account and the boundary conditions
calculating the single differential and total cross section are studied. The
calculated results well agree with the corresponding experimental data which
indicating the LDCS 1.0 program is good. It is also turned out that the
effect of tauon mass is not negligible in the GeV energy level.
\end{abstract}


\end{frontmatter}


{\bf PROGRAM SUMMARY}

\begin{small}
\noindent
{\em Manuscript Title:} On the lepton-nucleon neutral and charged current
                        deep inelastic scattering cross sections        \\
{\em Authors:} Xing-Long Li, Yu-liang Yan, Xiao-Mei Li, Dai-Mei Zhou, Xu Cai,
               Ben-Hao Sa        \\
{\em Program Title:} LDCS version 1.0        \\
{\em Journal Reference:}                     \\
{\em Catalogue identifier:}                  \\
{\em Licensing provisions:} none             \\
{\em Programming language:} FORTRAN 90                                   \\
{\em Computer:} computers with a FORTRAN 90 compiler                     \\
{\em Operating system:} Unix/Linux                                       \\
{\em RAM:} 128MB                                              \\
{\em Number of processors used:} 1                             \\
{\em Supplementary material:}                               
{\em Classification:} 13.60.Hb                 \\
{\em External routines/libraries:} LHAPDF 5.9.1                     \\
{\em Subprograms used:}                                       \\
{\em Catalogue identifier of previous version:}*              \\
{\em Journal reference of previous version:}*                  \\
{\em Does the new version supersede the previous version?:}*   \\
{\em Nature of problem:} \\
 The total cross section of the lepton-nucleon neutral current and charged
current deep inelastic scattering (DIS) is required in the
lepton-nucleus DIS simulations (e.g. in the PACIAE 2.2 model \cite{zhou}) and
in the design of DIS experiments. The incident and scattered lepton masses are
usually neglected \cite{PPR2012sref} in the lepton-nucleon DIS calculations.
However, Ref. \cite{NEUTsref} has pointed out that the tauon mass can not be
neglected in the GeV energy level. \\
{\em Solution method:}\\
The mass of the incident and scattered lepton are considered in the
lepton-nucleon DIS double differential cross section
at leading order and in the integral region calculating the single differential
and total cross section. Correspondingly, a FORTRAN program LDCS 1.0 is
constructed referring to the HERAfitter-1.0.0 \cite{HERAFITTERsref}.
The LHAPDF 5.9.1 parton distribution function (PDF) set \cite{LHAPDFsref} is
used.
   \\
{\em Reasons for the new version:}*\\
   \\
{\em Summary of revisions:}*\\
   \\
{\em Restrictions:}\\
 LDCS 1.0 can only be used calculating the unpolarized charged lepton-
unpolarized nucleon and the neutrino-unpolarized nucleon DIS double
differential, single differential, and total cross sections at leading order.
   \\
{\em Unusual features:}\\
 Since the lepton mass is included, the LDCS 1.0 program is able to consider
all 12 kinds of lepton-nucleon DIS, provided the corresponding PDF set is
at hand.
   \\
{\em Additional comments:}\\
   \\
{\em Running time:}\\
 Calculating the unpolarized electron-unpolarized proton neutral current DIS
total cross section at $\sqrt s$=318.7 GeV with HERAPDF1.5 LO PDF set
\cite{HERAPDFsref} takes 3.4 second. Calculating the neutrino-unpolarized
iron charged current DIS total cross section at $\sqrt s = 20.0$ GeV with
HKNlo PDF set \cite{HKNLOsref} takes 0.46 second.
\\

\end{small}

\section {Introduction}
The lepton-nucleon deep inelastic scattering (DIS) has been one of the most
important experiments in the high energy physics. It plays crucial role in the
investigation of the electroweak and strong
interactions (Quantum Chromodynamics, QCD) as well as the hadronic structure.

On the other hand, the lepton-nucleon (lepton-nucleus) DIS is significant in
the studies of time evolution of the hadronization and final hadronic state
(final hadronic rescattering). In the lepton-nucleus DIS simulations the
lepton-nucleon DIS total cross section is required in order to decide which
nucleon among the nucleons distributed randomly in the target nucleus sphere,
can collide with the incident lepton and how that DIS is evolved in the
nuclear medium.

The messages of lepton-nucleon DIS cross section are always embadded in the
complex parton distribution function(PDF) fitting packages, such as HERAfitter
\cite{HERAFITTERref1}. However, a simple but self-consistent program
calculating the lepton-nucleon DIS differential and total cross sections based
on the experimentally fitted PDF set is of benefit to the lepton-nucleus DIS
simulations.

To the end, we are devoted to construct a simple but self-consistent program
calculating the lepton-nucleon neutral current (NC) and charged
current (CC) DIS cross sections at leading order based on the fitted PDF.
As new features, the mass of the incident and scattered leptons are
taken into account and all 12 kinds of leptons are covered in the constructed
program LDCS 1.0.

The cross sections can only be calculated provided the PDF is knowledged.
Unfortunately, the PDF cannot be calculated in first principle and just can
be parameterized via fitting the measured lepton-nucleon differential cross
section to the corresponding theoretical calculation iteratively. For
example, the HERA1 and ZEUS groups measured the electron-proton DIS cross
sections at DESY \cite{HERAFITTERref1}, these data are then used to extract
the PDF set of HERAPDF 1.5 LO \cite{HERAPDFref} by HERAfitter program
\cite{HERAFITTERref1,HERAFITTERref2}.

In the section~\ref{theory}, we briefly introduce the theory of
lepton-nucleon NC and CC DIS at leading order. We compare the
calculated differential and total cross section to the corresponding
experimental data in the section~\ref{results}. Here the LDCS 1.0 is employed
calculating the unpolarized charged lepton-unpolarized proton DIS cross
sections based on the HERAPDF 1.5 LO PDF set \cite{HERAPDFref} and the
neutrino-unpolarized iron DIS cross sections based on the HKNlo iron PDF set
\cite{HKNLOref}. A brief summary is given in the section
\ref{summary}. The LDCS 1.0 program is described briefly in \ref{program}.
\section {Neutral and charged current cross section}
\label{theory}
\begin{figure}[htb]
  \centering
  \includegraphics[width=0.35\textwidth]{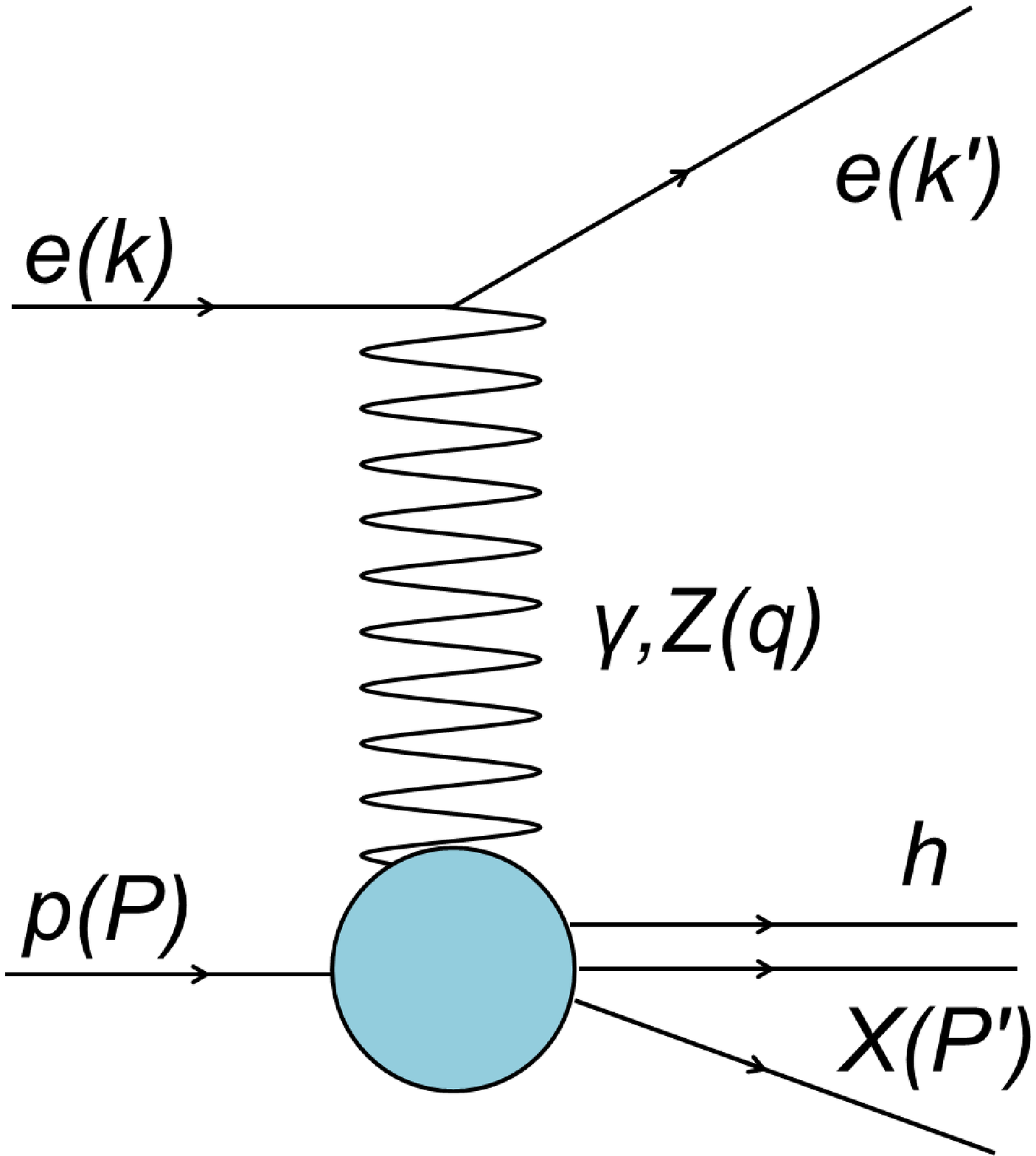}
  \includegraphics[width=0.35\textwidth]{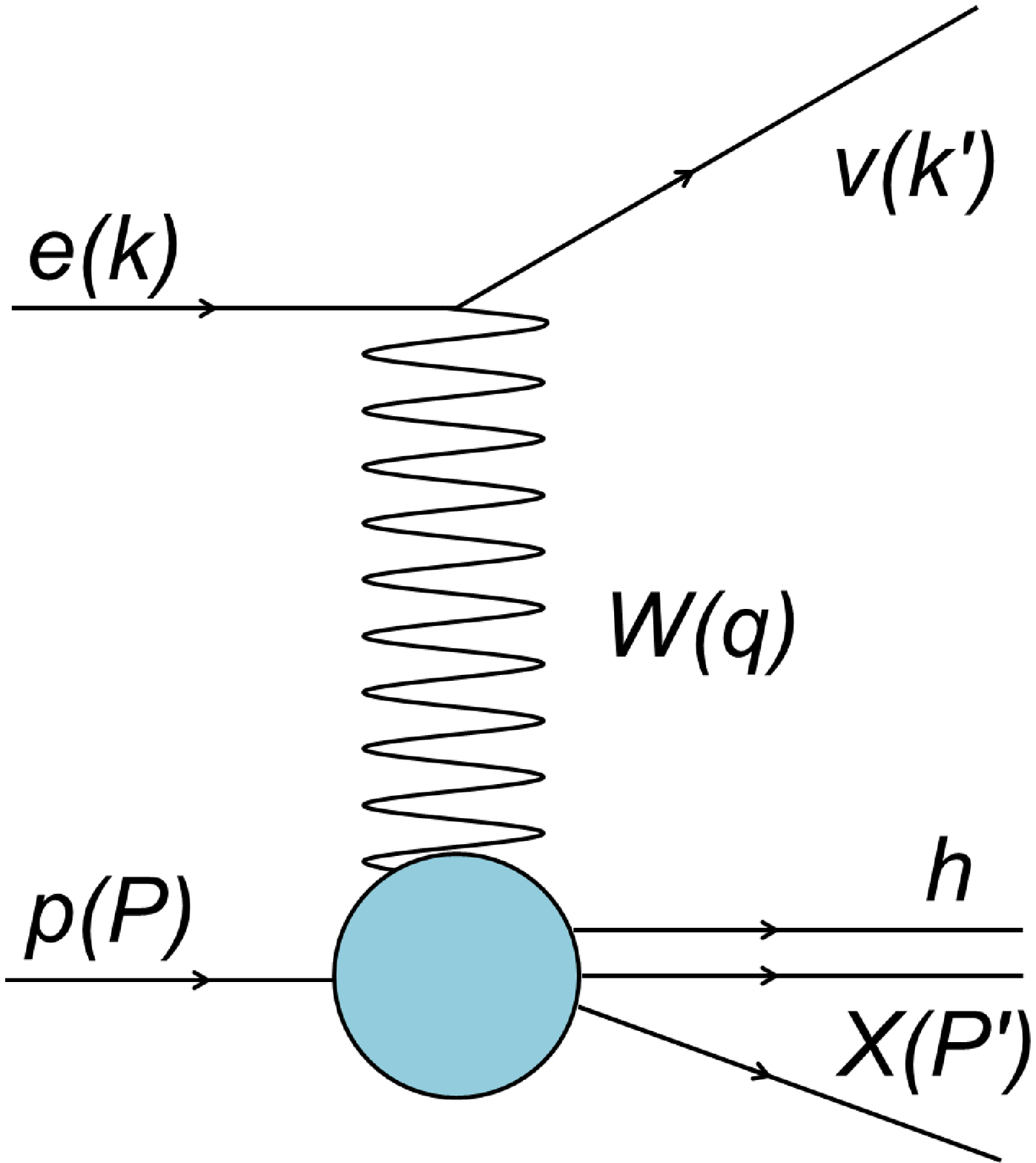}
  \caption{Feynman diagram of the electron-proton NC (\(ep \to eX\), left
panel) and CC (\(ep \to \nu X\), right panel) at leading order (Born
approximation).}
  \label{epLO}
\end{figure}

Figure~\ref{epLO} gives the Feynman diagram for the electron-proton NC (left
panel) and CC (right panel) DIS at leading order (Born approximation).
Correspondingly, the Kinematic variables defined in the nucleon rest frame
(adopted later on) is given in Tab.~\ref{variables}. We note that DIS is
restricted to the lepton-nucleon inelastic scattering process with
\({{Q^2}\gg{M^2}}\) and \({{W^2}\gg{M^2}}\).
\begin{table}[htb]
  \centering
  \caption{Kinematic variables in the lepton-nucleon DIS.}
  \begin{tabular}{lp{0.5\textwidth}}
  \hline
  \hline
  \(k = \left( {E_i,\vec k} \right),k' = \left( {E',\vec k'} \right)\) &
   4-momentum of the incident, scattered lepton \\
  \(m_i, m_o\) & Static mass of the incident, scattered lepton\\
  \(P = \left( {M,\vec 0 } \right)\) & 4-momentum of the nucleon\\
  \(q = k - k'\) & Squared 4-momentum transfer \\
  \(\nu  = \frac{{P \cdot q}}{M} = E_i - E'\) & Energy transfer to the
  target nucleon \\
  \(y = \frac{{P \cdot q}}{{P \cdot k}} = \frac{v}{E_i}\) & Fraction of the
   lepton energy lost\\
  \(x = \frac{{{Q^2}}}{{2P \cdot q}} = \frac{{{Q^2}}}{{2M \cdot \nu }}\)
  & Bjorken scaling variable\\
  \({W^2} = {\left( {P + q} \right)^2}=M^2+2ME_i y\left(1-x\right)\) &
  Squared invariant mass of the system X\\
  \({Q^2} =  - {q^2} = 2MxyE_i\) & Negative squared 4-momentum transfer\\
  \(s=\left(k+P\right)^2=\frac{Q^2}{xy}+M^2+m_i^2\) & Squared center-of-mass
   energy\\
  \hline
  \hline
  \label{variables}
\end{tabular}
\end{table}

In the lowest-order perturbative theory, the cross section of lepton-nucleon
DIS can be expressed as
\begin{equation}\label{origin}
\frac{{{\rm{d^2}}\sigma }}{{{\rm d}x{\rm d}y}} = \frac{{2\pi y{\alpha ^2}}}
{{{Q^4}}}\sum\limits_j {{\eta _j}L_j^{\mu \nu }} W_{\mu \nu }^j
\end{equation}
\cite{PPR2012ref}, where \(j=\rm{\gamma,Z,\gamma Z}\) for NC and
 \(j=\rm{W}^{\pm}\) for
CC. \(\alpha\) is the fine-structure constant. \({\eta _j}\) is defined by
\begin{eqnarray}\label{eta}
{\eta _\gamma } = 1 &,&  {\eta _{\gamma Z}} = \left( {\frac{{{G_F}M_Z^2}}
{{2\sqrt 2 \pi \alpha }}} \right)\left( {\frac{{{Q^2}}}{{{Q^2} + M_Z^2}}}
\right),
\nonumber \\
{\eta _Z} = \eta _{\gamma Z}^2 &,& {\eta _W} = \frac{1}{2}{\left( {\frac
{{{G_F}M_W^2}}{{4\pi \alpha }}\frac{{{Q^2}}}{{{Q^2} + M_W^2}}} \right)^2}
\end{eqnarray}
\cite{PPR2012ref}, where \(M_Z\) and \(M_W\) are, respectively, the mass of
\(Z\) and \(W\), and \(G_F\) is the Fermi constant given by
\begin{equation}\label{GF}
{G_F} = \frac{{{e^2}}}{{4\sqrt 2 {{\sin }^2}{\theta _W}M_W^2}}
\end{equation}

The lepton tensor \(L^{\mu\nu}\) is associated with the coupling of the
exchange boson to the lepton. For charged lepton (\(e=\pm 1\), helicity
\(\lambda=\pm 1\) \cite{PPR2012ref}), \(L^{\mu\nu}\) is given by
\begin{eqnarray}\label{cL}
&& L_\gamma ^{\mu \nu } = 2\left( {{k^\mu }k{'^\nu } + k{'^\mu }{k^\nu }
 - \left( {k \cdot k' - m_i^2} \right){g^{\mu \nu }} - i\lambda
 {\varepsilon ^{\mu \nu \alpha \beta }}{k_\alpha }k{'_\beta }} \right),
 \nonumber \\
&& L_{\gamma Z}^{\mu \nu } = \left( {g_V^{cl} + e\lambda g_A^{cl}} \right)
L_\gamma ^{\mu \nu },\nonumber \\
&& L_Z^{\mu \nu } = {\left( {g_V^{cl} + e\lambda g_A^{cl}} \right)^2}
L_\gamma ^{\mu \nu },\nonumber \\
&& L_W^{\mu \nu } = {\left( {1 + e\lambda } \right)^2}L_\gamma ^{\mu \nu }
\end{eqnarray}
where \(cl\) refers to charged lepton.
\(g_V^{cl} = -1/2 + 2{\sin ^2}{\theta _W}\) is
the weak vector coupling of the charged lepton to Z. \(g_A^{cl} =  -1/2\) is
the weak axial-vector coupling of the charged lepton to Z. \(\theta_W\) is the
Weinberg angle. \(g^{\mu \nu }\) is the metric tensor and
\(\varepsilon ^{\mu \nu \alpha \beta }\) is the completely antisymmetric unit
 tensor. Because the number of left-hand ($\lambda=-1$) charged lepton is
 equal to the right-hand ($\lambda=1$) charged lepton, the unpolarized charged
  lepton-unpolarized nucleon DIS cross section is just the average of the
  left- and right-hand contributions.

For neutrino (\(e=0\), \(\lambda=\pm 1\)), because \(\gamma\) exchange does
not exist and only the left-handed neutrino (\(\lambda =-1\)) as well as the
right-handed anti-neutrino (\(\lambda=+1\)) are observed so far, we have
\begin{eqnarray}\label{vL}
&& L_Z^{\mu \nu } = 2{\left( {g_V^{nu} + g_A^{nu}} \right)^2}
\left( {{k^\mu }k{'^\nu } + k{'^\mu }{k^\nu } - \left( {k \cdot k' - m_i^2}
 \right){g^{\mu \nu }} - i\lambda {\varepsilon ^{\mu \nu \alpha \beta }}
 {k_\alpha }k{'_\beta }} \right),\nonumber \\
&& L_W^{\mu \nu } = 8\left( {{k^\mu }k{'^\nu } + k{'^\mu }{k^\nu } -
\left( {k \cdot k' - m_i^2} \right){g^{\mu \nu }} - i\lambda
{\varepsilon ^{\mu \nu \alpha \beta }}{k_\alpha }k{'_\beta }} \right)
\end{eqnarray}
where \(g_V^{nu} =\frac{1}{2}\), \(g_A^{nu} =\frac{1}{2}\), and the
superscript \(nu\) refers to neutrino. One has to point out here that the
efforts searching for right-handed neutrino are in progress \cite{rightneu},
provided it is found eventually then the equation (\ref{vL}), of
course, should be changed.

Neglecting CP-violating effect, the nucleon tensor \(W_{\mu \nu }\) can be
expressed as \\
\begin{eqnarray}\label{Wuv}
&& {W_{\mu \nu }} = \left( { - {g_{\mu \nu }} + \frac{{{q_\mu }{q_\nu }}}
{{{q^2}}}} \right){F_1}\left( {x,{Q^2}} \right) + \frac{{{{\hat P}_\mu }
{{\hat P}_\nu }}}{{P \cdot q}}{F_2}\left( {x,{Q^2}} \right) \nonumber\\
&& + i{\varepsilon _{\mu \nu \alpha \beta }}\frac{{{q^\alpha }{P^\beta }}}
{{2P \cdot q}}{F_3}\left( {x,{Q^2}} \right) -i{\varepsilon _
{\mu \nu \alpha \beta }}\frac{{{q^\alpha }{s^\beta }}}{{P \cdot q}}{g_1}
\left( {x,{Q^2}} \right) \nonumber\\
&& -i{\varepsilon _{\mu \nu \alpha \beta }} \frac{q^\alpha
 \left({P\cdot qs^\beta - s\cdot qP^\beta}\right)}{\left({P\cdot q}\right)^2}
 {g_2}\left( {x,{Q^2}} \right) \nonumber\\
&& + \frac{1}{P\cdot q}\left({\frac{1}{2}\left({{{\hat P}_\mu }{{\hat s}_\nu }
+{{\hat P}_\nu }{{\hat s}_\mu }}\right)-\frac{s\cdot q}{P\cdot q}{{\hat P}_\mu }
{{\hat P}_\nu }}\right){g_3}\left( {x,{Q^2}} \right) \nonumber\\
&& +\frac{s\cdot q}{P\cdot q}\left({\frac{{{\hat P}_\mu }{{\hat P}_\nu }}
{P\cdot q}g_4\left({x,Q^2}\right)+\left({-g_{\mu \nu} +\frac{q_\mu q_\nu}
{q^2}}\right)g_5\left({x,Q^2}\right)}\right)
\end{eqnarray}
\cite{PPR2012ref, WSFref, JXSFref}. Here \(q^\mu\) is the 4-momentum transfer,
\(P^\mu\) and \(s^\mu\) are the 4-momentum and spin of nucleon, respectively.
\(\hat P_\mu ,\hat s_\mu\) are defined as follows
\begin{equation}\label{Pshat}
{\hat P_\mu } = {P_\mu } - \frac{{P \cdot q}}{{{q^2}}}{q_\mu },
\quad{     } {\hat s_\mu } = {s_\mu } - \frac{{s \cdot q}}{{{q^2}}}{q_\mu }.
\end{equation}
\({F_i}\left( {x,{Q^2}} \right)\left({i=1,2,3}\right)\) are unpolarized
structure functions, and
 \({g_i}\left({x,{Q^2}} \right)\left({i=1,2,3,4,5}\right)\) are polarized
structure functions \cite{WSFref, JXSFref}.

For the lepton-unpolarized nucleon DIS, the nucleon tensor can be simplified
as
\begin{eqnarray}\label{uWuv}
&& {W_{\mu \nu }} = \left( { - {g_{\mu \nu }} + \frac{{{q_\mu }{q_\nu }}}
{{{q^2}}}} \right){F_1}\left( {x,{Q^2}} \right) + \frac{{{{\hat P}_\mu }
{{\hat P}_\nu }}}{{P \cdot q}}{F_2}\left( {x,{Q^2}} \right) \nonumber\\
&& + i{\varepsilon _{\mu \nu \alpha \beta }}\frac{{{q^\alpha }{P^\beta }}}
{{2P \cdot q}}{F_3}\left( {x,{Q^2}} \right)
\end{eqnarray}
In the naive quark-parton model \cite{quark-partonref}, \(F_1\) and \(F_2\)
 are approximately related by
Callan-Gross relation
\begin{equation}\label{CG}
F_1=\frac{1}{2x}\left({1+4\frac{x^2M^2}{Q^2}}\right)F_2.
\end{equation}
Inserting Eqs.~(\ref{eta}), (\ref{cL}), (\ref{vL}) and (\ref{uWuv}) into
Eq.~(\ref{origin}) we obtain
\begin{eqnarray}\label{DIS}
&& \frac{{\rm{d^2}} \sigma^{I}}{{\rm{d}} x {\rm{d}} y}=\frac{8\pi
\alpha ^2 ME_i}{Q^4}\left( {c_1F _1^I+c_2F_2^I+c_3xF_3^I}\right),\nonumber\\
&& c_1=xy^2-\frac{\left({{m_i}^2-{m_o}^2}\right)^2}{8xM^2E^2_i}-\frac
{y\left({5{m_i}^2-{m_o}^2}\right)}{4ME_i}\nonumber,\\
&& c_2=1-y \nonumber \\
&& +\frac{\left({{m_i}^2-{m_o}^2}\right)\left({4x^2M^2+{m_i}^2-{m_o}^2}\right)}
{16x^2M^2{E_i}^2}-\frac{\left({{m_i}^2-{m_o}^2}\right)\left({y-4}\right)
+4M^2x^2y}{8xME_i}\nonumber,\\
&& c_3=\frac{\lambda y(y-2)}{2}-\frac{\lambda y\left({{m_i}^2-{m_o}^2}\right)}
{4xME_i}
\end{eqnarray}
\begin{eqnarray}\label{allI}
&& F_2^{clNC} = F_2^{\gamma}-\left({g_V^{cl}+e\lambda g_A^{cl}}\right)\eta_
{\gamma Z}F_2^{\gamma Z}+\left( {g_V^{cl} + e\lambda g_A^{cl}} \right)^2
\eta_ZF_2^Z, \nonumber\\
&& xF_3^{clNC} = -\left({g_V^{cl}+e\lambda g_A^{cl}}\right)\eta_{\gamma Z}
xF_3^{\gamma Z}+\left( {g_V^{cl} + e\lambda g_A^{cl}} \right)^2\eta_ZxF_3^Z,
\nonumber\\
&& F_2^{clCC} = \left({1+e\lambda}\right)^2\eta_WF_2^W, \nonumber\\
&& xF_3^{clCC} = \left({1+e\lambda}\right)^2\eta_WxF_3^W,\nonumber\\
&& F_2^{nuNC} = \left( {g_V^{nu} + g_A^{nu}} \right)^2\eta_ZF_2^Z, \nonumber\\
&& xF_3^{nuNC} = \left( {g_V^{nu} + g_A^{nu}} \right)^2\eta_ZxF_3^Z,\nonumber\\
&& F_2^{nuCC} = 4\eta_WF_2^W, \nonumber\\
&& xF_3^{nuCC} = 4\eta_WxF_3^W,
\end{eqnarray}
for the lepton-unpolarized nucleon DIS. In the Eq.~(\ref{DIS})
\(I=clNC,nuNC,clCC$, and $nuCC\) (\(cl\)=\(e^\pm, \mu^\pm, \tau^\pm\); \(nu\)
=\(\nu_e, \overline{\nu}_e, \nu_{\mu}, \overline{\nu}_{\mu}, \nu_{\tau},
\overline{\nu}_{\tau}\)).

In the quark-parton model, the structure functions are related to the parton
distribution function (\(q(x,Q^2)\)). For neutral current DIS, we have
\begin{eqnarray}\label{NCF23}
{[F_2^\gamma ,F_2^{\gamma Z},F_2^Z]} &=& x\sum\limits_q{[e_q^2,2e_qg_V^q,
{g_V^q}^2+{g_A^q}^2](q+\overline q)},\nonumber\\
{[xF_3^\gamma ,xF_3^{\gamma Z},xF_3^Z]} &=& x\sum\limits_q{[0,2e_qg_A^q,
2g_V^qg_A^q](q-\overline q)}
\end{eqnarray}
where \(e^q\) is the charge of \(q\) quark. \(g_V^q\) as well as \(g_A^q\)
are, respectively, the weak vector coupling between \(q\) and \(Z\) as well
as the weak axial-vector coupling between \(q\) and \(Z\). They are
\begin{eqnarray}\label{gqva}
q=u,c,t: & g_A^q=\frac{1}{2}, & g_V^q=g_A^q-\frac{4}{3}sin^2\theta _W;\nonumber\\
q=d,s,b: & g_A^q=-\frac{1}{2}, & g_V^q=g_A^q+\frac{2}{3}sin^2\theta _W.
\end{eqnarray}
For charged current DIS, we have
\begin{eqnarray}\label{Fw-}
F_2^W=2x\left({u+\overline d+c+\overline s+t+\overline b}\right),\nonumber\\
xF_3^W=2x\left({u-\overline d+c-\overline s+t-\overline b}\right)
\end{eqnarray}
for incident lepton of \(e^-, \mu^-, \tau^-, \overline{\nu}_e, \overline{\nu}_\mu\)
and $\overline{\nu}_\tau\), as well as
\begin{eqnarray}\label{Fw+}
F_2^W=2x\left({\overline u+d+\overline c+s+\overline t+b}\right),\nonumber\\
xF_3^W=2x\left({-\overline u+d-\overline c+s-\overline t+b}\right)
\end{eqnarray}
for incident lepton of \(e^+, \mu ^+, \tau ^+ \), \(\nu_e, \nu_{\mu}\), and
\(\nu_{\tau}\).

In the lepton-nucleon DIS, because of \(Q^2 \gg m_i^2\) and \(Q^2 \gg m_o^2\),
one may assume \(m_i^2=m_o^2=0\). Therefore Eq.~(\ref{DIS}) can be
simplified to
\begin{equation}\label{epDIS}
\frac{{\rm{d^2}} \sigma^{I}}{{\rm{d}} x {\rm{d}} y}=\frac{4\pi \alpha ^2}{xyQ^2}
\left(xy^2F_1^I+\left(1-y-\frac{x^2y^2M^2}{Q^2}\right)F_2^I-\lambda
\left(y-\frac{y^2}{2}\right)xF_3^I\right).
\end{equation}
Note, a relation of
\begin{equation}\label{trans}
\frac{{\rm d^2}\sigma}{{\rm d}x{\rm d}Q^2}=\frac{y}{Q^2}\frac
{{\rm d^2}\sigma}{{\rm d}x{\rm d}y}
\end{equation}
is always required in the argument transformation from $dxdy$ to $\frac{Q^2}
{y}dxdQ^2$.

The total cross section can be expressed by the differential cross section
(Eq.~(\ref{DIS})) as follows
\begin{equation}\label{integ}
\sigma=\int_{x_{min}}^{x_{max}}\int_{y_{min}}^{y_{max}}\frac{{\rm d^2}
\sigma}{{\rm d}x{\rm d}y}{\rm d}x{\rm d}y
\end{equation}
where the integral limits should be decided by the scattering kinematics. In
order that the direction of incident lepton is set on the $z$ axis and the
lepton scattering angel is denoted as \(\theta\), one then has
\begin{eqnarray}\label{fortheta}
&& k=\left({E_i,0,0,\sqrt{E_i^2-m_i^2}}\right),\nonumber\\
&& k'=\left({E',0,\sqrt{E'^2-m_o^2}sin\theta,\sqrt{E'^2-m_o^2}cos\theta}
 \right),\nonumber\\
&& 2ME_ixy=Q^2=-{\left({k-k'}\right)}^2,\nonumber\\
&& y=\frac{E_i-E'}{E_i}.
\end{eqnarray}
Further one can deduce
\begin{equation}\label{cos2th}
cos^2\theta=\frac{\left({m_i^2+m_o^2+2E_i^2(y-1)+2E_iMxy}\right)^2}
{4\left({E_i^2-m_i^2}\right) \left({E_i^2\left({y-1}\right)^2-m_o^2}\right)}
\end{equation}
from Eq.~(\ref{fortheta}). Together with the facts of \(cos^2\theta\le 1\),
$0\le x\le1$ and $0\le y\le1$, the $x$ and $y$ kinematic limits are obtained
\begin{eqnarray}\label{xykl}
&& max\left(0, \frac{m_o^2-m_i^2}{2M\left({E_i-m_o}\right)}\right) \le x \le 1
,\nonumber\\
&& A-B \le y \le A+B, \nonumber\\
&& A=\frac{2MxE_i^2+E_i\left({m_i^2-m_o^2}\right)-Mx\left({m_i^2+m_o^2}
\right)} {2E_i\left({M^2x^2+2E_iMx+m_i^2}\right)}, \nonumber\\
&& B=\frac{\sqrt{\left({E_i^2-m_i^2}\right)\left({4M^2x^2\left({E_i^2-m_o^2}
\right)+4E_iMx\left({m_i^2-m_o^2}\right)+\left({m_i^2-m_o^2}\right)^2}\right)}}
{2E_i\left({M^2x^2+2E_iMx+m_i^2}\right)}. \nonumber\\
\end{eqnarray}

Meanwhile, the $x$ and $y$ integral limits must keep with the DIS characters
of \({{Q^2}\gg {M^2}}\) and \({{W^2}\gg{M^2}}\), which lead to
\begin{equation}\label{xminmax}
\frac{Q_{min}^2}{2ME_i} \le x \le 1-\frac{W_{min}^2-M^2}{2ME_i},
\end{equation}
and
\begin{equation}\label{yminmax}
max\left(\frac{Q_{min}^2}{2ME_i x}, \frac{W_{min}^2-M^2}{2ME_i
\left(1-x\right)} \right) \le y
\end{equation}
where \(0\le y \le 1\) is employed.

Combining Eq.~(\ref{xykl}), (\ref{xminmax}) and (\ref{yminmax}),
 we obtain $x$ and $y$ integral limits eventually
\begin{equation}\label{xscope}
max\left(\frac{m_o^2-m_i^2}{2M\left({E_i-m_o}\right)},
\frac{Q_{min}^2}{2ME_i}\right) \le x \le 1-\frac{W_{min}^2-M^2}{2ME_i},
\end{equation}
\begin{equation}\label{yscope}
max\left(A-B, \frac{Q_{min}^2}{2ME_i x}, \frac{W_{min}^2-M^2}
{2ME_i \left(1-x\right)} \right) \le y \le A+B.
\end{equation}
\section {Results}
\label{results}
Hereafter, we will mainly use the LDCS 1.0 program (the equation~(\ref{DIS}))
to calculate the cross sections for the various lepton-nucleon DIS and compared
with the corresponding experimental data in order to prove the correction of
the LDCS 1.0 program.

The double differential cross sections of the unpolarized electron-unpolarized
proton DIS at $\sqrt{s}$=318.7 GeV are shown in Fig.~\ref{epdd} for different
$x$ values. In this figure the solid and open triangles are experimental data
measured by HERA1 and ZEUS \cite{HERAFITTERref1}, the solid and dashed lines
are theoretical results calculated with HERAPDF1.5 LO PDF set
\cite{HERAPDFref}. One sees in this figure that the experimental data are
well reproduced by the theoretical calculations.
\newpage
\begin{center}
  \includegraphics[width=0.8\textwidth]{ddNC.eps}
  \\[0.02\textwidth]
  \includegraphics[width=0.8\textwidth]{ddCC.eps}
  \figcaption{Experimental and calculated double differential cross sections
of the unpolarized electron-unpolarized proton DIS at $\sqrt{s}$=318.7 GeV.
Solid and dashed lines
are calculated with HERAPDF1.5 LO PDF set \cite{HERAPDFref}. Solid and open
triangles with error bars are experimental data measured by HERA1 and ZEUS
\cite{HERAFITTERref1}. }
  \label{epdd}
\end{center}

We give the theoretical single differential cross sections \(d\sigma/dx\)
(solid curves) obtained by integrating the corresponding theoretical double
differential cross sections (as shown in Fig~\ref{epdd}) over $Q^2$ with
\(Q^2_{min}=$1000 GeV$^2\) and \(y<0.9\) in Fig.~\ref{epdx}. In this figure
the solid and open circles with error bars are the corresponding experimental
data taken from \cite{DESY00187ref,DESY03038ref}. One sees in this figure
again that the theory well agrees with the experiment.

\begin{center}
  \includegraphics[width=0.45\textwidth]{e-pNC.eps}
  \includegraphics[width=0.45\textwidth]{e+pNC.eps}
  \\[0.02\textwidth]
  \includegraphics[width=0.45\textwidth]{e-pCC.eps}
  \includegraphics[width=0.45\textwidth]{e+pCC.eps}
  \figcaption{The experimental and calculated single differential cross
sections \(d\sigma/dx\) in the unpolarized electron-unpolarized proton DIS
at \(\sqrt{s}$=318.7 GeV.
Solid lines are calculated with HERAPDF1.5 LO PDF set \cite{HERAPDFref}.
Solid and open circles with error bars are experimental data taken from
\cite{DESY00187ref,DESY03038ref}. }
  \label{epdx}
\end{center}

The unpolarized electron-unpolarized proton DIS total cross section
\(\sigma _{DIS}\) (NC+CC) as a function of $\sqrt s$ calculated with
HERAPDF1.5 LO PDF set \cite{HERAPDFref} and kinematical
cuts of \(Q^2_{min}=$1.0 GeV$^2\) and \(W^2_{min}=$1.96 GeV$^2\) is shown in
Fig.~\ref{eptotal} as black solid line. In this figure the red and blue circles
are, respectively, the pp and $\gamma$p total cross sections taken from
\cite{PPR2012ref}. We see in Fig~\ref{eptotal} that the $e^-p$ DIS total cross
section is a few order of magnitude smaller than pp at the range of
$\sqrt s\leq$10$^4$ GeV.

\begin{center}
  \includegraphics[width=0.8\textwidth]{total.eps}
  \figcaption{The calculated unpolarized electron-unpolarized proton DIS total
  cross section and the pp as well as \(\gamma p\) total cross sections
  (taken from \cite{PPR2012ref}). }
  \label{eptotal}
\end{center}
We compare the unpolarized electron-unpolarized proton DIS total cross section
(solid line) with the corresponding \(\mu^-p\) (crosses) and \(\tau^-p\) (dashed
line) DIS total cross sections in the Fig.~\ref{eutDIS}. The difference among
them stems from their mass.
\begin{center}
  \includegraphics[width=0.8\textwidth]{totaleut.eps}
  \figcaption{The total cross sections of unpolarized electron-unpolarized
  proton, and the corresponding \(\mu^-p\) as well as \(\tau^-p\) DIS
  calculated with HERAPDF1.5 LO PDF set \cite{HERAPDFref} and kinematic
  cuts of \(Q^2_{min}=\) 1.0 GeV$^2$ as well as \(W^2_{min}=\)1.96 GeV$^2$. }
  \label{eutDIS}
\end{center}

The solid circles and triangles as well as open circles and triangle in
Fig.~\ref{vfe} present, respectively, the \(\nu_{\mu}Fe\) as well as
\(\bar{\nu}_{\mu}Fe\) (here $Fe$ is unpolarized) CC inclusive
scattering total cross sections experimental data ($\sigma_{cc}^{exp1}$ as
well as $\sigma_{cc}^{exp2}$). The triangle data are
taken from \cite{MINOSref} and the circle data from \cite{SELIGMANref}. As
mentioned in \cite{MINOSref} that their \(\nu_{\mu}Fe\) and \(\bar{\nu}_
{\mu}Fe\) CC inclusive scattering cross sections were measured by the
iron-scintillator detector with a 6.1\% excess of neutron over proton. This
was corrected by the NEUGEN3 cross section method \cite{neugen3}. The
correction is about -2\% for neutrino and +2\% for antineutrino. Therefore,
the expected cross section of $\nu_{\mu}$ ($\bar\nu_{\mu}$) CC inclusive
scattering off the isoscalar target is (1-0.02)$\sigma_{cc}^{exp1}$
((1+0.02)$\sigma_{cc}^{exp2}$). This cross section is quoted as the cross
section of $\nu_{\mu}N$ ($\bar\nu_{\mu}N$) CC inclusive scattering in
\cite{PPR2012ref}, $N$ here is refers to the nucleon (neutron
and/or proton).

The dashed and solid curves in Fig.~\ref{vfe} are the theoretical total cross
sections of $\nu_{\mu}Fe$ and $\bar{\nu}_{\mu}Fe$ DIS calculated by
Eq.~(\ref{DIS}) and with the iron PDF set (taken from
\cite{HKNLOref}), the kinematic cuts of $Q_{min}^2$=1.0 GeV$^2$ and
$W_{min}^2$=1.9 GeV$^2$, as well as the correction factor of 0.98 for
neutrino and 1.02 for antineutrino. One sees in Fig.~\ref{vfe} that the
theory agrees with experiment fairly well at the high incident lepton energy
region ($E_i\geq$100 GeV), but the agreement is bad at the low energy region.
This is because the theory is for the DIS process only but the experiment is
for the inclusive scattering which is a combination of the quasi-elastic
scattering, DIS, and resonance production. And the DIS is dominated at the
high incident lepton energy region but at the low energy region the
quasi-elastic scattering and resonance production are dominated
\cite{PPR2012ref}. Therefore, one may conclude either that the equation
(\ref{DIS}) (the LDCS 1.0 program) is also work for the neutrino induced DIS.

\begin{center}
  \includegraphics[width=0.8\textwidth]{vfe.eps}
  \figcaption{The ratio of total cross section to the incident energy. The
solid and open triangles as well as the solid and open circles are
experimental data of \(\nu _{\mu}Fe\) and \(\bar{\nu} _{\mu}Fe\) CC inclusive
scattering taken from \cite{MINOSref} as well as \cite{SELIGMANref},
respectively. The solid and dashed curves are theoretical results of
\(\nu _{\mu}Fe\) and \(\bar{\nu} _{\mu}Fe\) DIS calculated by Eq.~(\ref{DIS})
with the kinematical cuts of \(Q_{min}^2=1.0\) GeV\(^2\) and \(W_{min}^2=
1.96\) \(GeV^2\) as well as the iron PDF set \cite{HKNLOref}.}
  \label{vfe}
\end{center}

Fig.~\ref{comp} shows the lepton mass effect on the NC (panel (a)) and CC
((b)) unpolarized charged lepton-unpolarized proton DIS total
cross sections as well as on the NC ((c)) and CC ((d)) neutrino-unpolarized
iron DIS total cross sections. The proton PDF used in the panels (a) and
(b) calculations is taken from \cite{HERAPDFref} and the iron PDF used in
the panels (c) and (d) calculations is taken from
\cite{HKNLOref}. We see in Fig.~\ref{comp} that the mass effect on the NC
lepton-proton and neutrino-iron DIS total cross section is weak
(cf. panels (a) and (c)). The effect on the CC lepton-proton DIS total cross
section is also weak(cf. panel (b)) but on the CC neutrino-iron DIS is
visible(cf. panel (d)).
\newpage
\begin{center}
  \includegraphics[width=0.45\textwidth]{clNC.eps}
  \includegraphics[width=0.45\textwidth]{clCC.eps}
  \\[0.02\textwidth]
  \includegraphics[width=0.45\textwidth]{vNC.eps}
  \includegraphics[width=0.45\textwidth]{vCC.eps}
  \figcaption{The lepton mass effect on the NC and CC unpolarized charged
lepton-unpolarized proton DIS as well as on the NC and CC neutrino-unpolarized
iron DIS. In the calculations the kinematic
cuts are: \(Q_{min}^2=1.0$ GeV$^2$ and $W_{min}^2=1.96$ GeV\(^2\).}
  \label{comp}
\end{center}

Fig.~\ref{bd1} shows the lepton mass effect on \(x$ and $y\) integral region
in the total cross section calculations for the unpolarized tauon-unpolarized
proton CC DIS at different $\sqrt s$. Here we see that the integral region
increases with increasing $\sqrt s$. We compare the lepton mass effect on the
different unpolarized charge lepton-unpolarized proton (neutrino-unpolarized
proton) NC and CC DIS at $\sqrt s$=3.5 GeV in the panels (a) and (c)((b) and
(d)) of Fig.~\ref{bd2}, respectively. One sees in the panels (a) and (c) ((b)
and (d)) of Fig.~\ref{bd2} that the larger tauon mass leads to that the
\(x$ and $y\) integral region in tauon induced DIS is smaller than the ones in
electron and/or muon induced DIS.
\begin{center}
  \includegraphics[width=0.8\textwidth]{15CC.eps}
  \figcaption{The \(x$ and $y\) integral region in unpolarized
  tauon-unpolarized proton DIS total cross section calculations at different
  $\sqrt s$. \(Q_{min}^2$=1.0 GeV$^2$ and $W_{min}^2$=1.96 GeV$^2\).}
  \label{bd1}
\end{center}
\newpage
\begin{center}
  \includegraphics[width=0.45\textwidth]{a.eps}
  \includegraphics[width=0.45\textwidth]{b.eps}
  \\[0.02\textwidth]
  \includegraphics[width=0.45\textwidth]{c.eps}
  \includegraphics[width=0.45\textwidth]{d.eps}
  \figcaption{The \(x$ and $y\) integral region in different lepton-proton DIS
total cross section calculations at $\sqrt{s}$=3.5 GeV. $Q_{min}^2$=1.0
GeV$^2$ and $W_{min}^2$=1.96 GeV$^2$ }
  \label{bd2}
\end{center}
\section {Summary}
\label{summary}

The Monte Carlo simulation is one of methods investigating the lepton-nucleon
and lepton-nucleus DIS. In the lepton-nucleus DIS, the nucleons in target
nucleus are first random distributed in the target nuclear sphere according
to the Woods-Saxon and 4$\pi$ isotropic distributions. The collision
possibility is then considered between the incident lepton and each one of the
target nucleons, here the lepton-nucleon DIS total cross section is
necessary.

Although, the message of lepton-nucleon DIS cross section are always embadded
in the complex PDF fitting packages such as HERAfitter \cite{HERAFITTERref1}.
However, for the benefit of lepton-nucleus DIS Monte Carlo simulation, a
simple but self-consistent program calculating the lepton-nucleon DIS
differential and total cross sections based on the experimentally fitted PDF
set is highly required.

Therefore, we pick up the concerned messages (subprograms) from the PDF
fitting packages and compose a simple but self-consistent program (LDCS 1.0)
for the calculation of the unpolarized charged lepton-unpolarized nucleon and
the neutrino-unpolarized nucleon DIS differential and total cross sections.
Before that, we have first briefly introduced the basic theory about the
lepton-nucleon NC and CC DIS. The equation of~(\ref{DIS}) is then
derived calculating the NC and CC differential cross sections for the
lepton-unpolarized nucleon DIS at leading order with the incident and
scattered lepton masses taken into account.

Then we have compared the calculated lepton-nucleon and/or lepton-iron
double, single differential, and total cross sections to the corresponding
experimental data. The good agreements between the theoretical results and
the corresponding experimental data indicate that the program LDCS 1.0 works
very well. Additionally, the integral region of Eq.~(\ref{xscope}) and
(\ref{yscope}) for the arguments of \(x\) and \(y\) are studied in detail.
The investigations in both the cross section and the integral region lead to
a conclusion that the mass of tauon can not be neglected in the GeV energy
level.

\appendix
\section{A brief description for the program}
\label{program}

The program LDCS 1.0, calculating the unpolarized charged lepton-unpolarized
nucleon and neutrino-unpolarized nucleon NC and CC DIS cross
sections, is mainly based on HERAfitter-1.0.0 program. The LHAPDF 5.9.1
packet is employed during running. The key subroutines in the program are
defined in `ddcs.f' and listed as follows:
\begin{itemize}
\item subroutine init\\
input: none\\
output: none\\
function: define the values of some constants
\item subroutine GetNCddsigma(x, Q2, y, NPT, M, lepin, xsec)\\
input: x, Q2 (i.e. \(Q^2\)), y, NPT (the size of array x, q2, y and xsec),
M, lepin (the number of incident lepton, 11: \(e^-\), -11: \(e^+\),
12: \(\nu _e\), -12: \(\bar{\nu}_e\), 13: \(\mu ^-\), -13: \(\mu ^+\),
14: \(\nu _{\mu}\), -14: \(\bar{\nu} _{\mu}\), 15: \(\tau ^-\),
-15: \(\tau ^+\), 16: \(\nu _{\tau}\), -16: \(\bar{\nu} _{\tau}\))\\
output: xsec (\(d^2\sigma _{NC}/dxdy\))\\
function: calculate the neutral current differential cross sections
\item subroutine GetCCddsigma(x, Q2, y, NPT, M, lepin, xsec)\\
input: x, Q2, y, NPT, M, lepin\\
output: xsec (\(d^2\sigma _{CC}/dxdy\))\\
function: calculate the charged current differential cross sections
\item subroutine getSF(SFD, x, Q2, SF, NPT)\\
input: x, Q2, NPT, SFD (coefficients of the linear combination to calculate
structure function from \(q\left(x,Q^2\right)\).)\\
output: SF (the values of structure function \(F_2\) or \(xF_3\))\\
function: calculate \(F_2\) or \(xF_3\) by LHAPDF 5.9.1 \cite{LHAPDFref}
\end{itemize}

In ``main.f'', the double differential cross sections (\(d^2\sigma /dxdy\)),
single differential cross sections (\(d\sigma /dx\)), and the total cross
section (\(\sigma_{DIS}\)) are calculated in subroutine of `ddsigma',
`dxsigma' and `sigma', respectively. The Simpson method is employed for the
integrations. ``SFdefine.inc'' defines the coefficients of the linear
combination to calculate structure function from quark distribution function
of \(q\left(x,Q^2\right)\). The contribution of top quark is neglected.
``couplings.inc'' gives the declarations for some constants. ``input.txt'' sets
the type and parameters for the calculation. ``output.txt'' saves the results.
There are four examples for the input and output files given in folder
``example''. ``readme.txt'' guides the program installation and running.

\end{document}